\documentclass[submitting]{nst}

\usepackage{subfigure,dcolumn}
\usepackage[T2A,T1]{fontenc}
\usepackage[russian,english]{babel}
\DeclareUnicodeCharacter{0301}{\'{e}}

\usepackage{listings}
\begin{document}

\title{Investigating  the elliptic anisotropy of identified particles in p--Pb collisions with a multi-phase transport model}\thanks{Supported by the Key Laboratory of Quark and Lepton Physics (MOE) in Central China Normal University (No.QLPL2022P01, QLPL202106), the national key research, development program of China (No.2018YFE0104700), National Natural Science Foundation of China (No.12175085), Fundamental research funds for the Central Universities (No.CCNU220N003), Natural Science Foundation of Hubei Provincial Education Department (No.Q20131603)}

\author{Si-Yu Tang}
   
    \affiliation{School of Mathematical and Physical Sciences, Wuhan Textile University, Wuhan 430200, China}
    
\author{Liang Zheng}
    \email[Correspondence email address: ]{zhengliang@cug.edu.cn}
    \affiliation{School of Mathematics and Physics, China University of Geosciences (Wuhan), Wuhan 430074, China}
\author{Xiao-Ming Zhang}
    \email[Correspondence email address: ]{xiaoming.zhang@ccnu.edu.cn}
    \affiliation{Key Laboratory of Quark and Lepton Physics (MOE) and Institute of Particle Physics, Central China Normal University, Wuhan 430079, China}
\author{Ren-Zhuo Wan}
    \email[Correspondence email address: ]{wanrz@wtu.edu.cn}
    \affiliation{Hubei Key Laboratory of Digital Textile Equipment, Wuhan Textile University, Wuhan 430200, China}
    \affiliation{School of Electronic and Electrical Engineering, Wuhan Textile University, Wuhan 430200, China}
\begin{abstract}
 The elliptic azimuthal anisotropy coefficient ($v_{2}$) of the identified particles at midrapidity ($|\eta|<0.8$) was investigated in p--Pb collisions at $\sqrt{s_\mathrm{NN}}=$ 5.02 TeV using a multi-phase transport model (AMPT). The calculations of differential $v_{2}$ based on the advanced flow extraction method of light flavor hadrons (pions, kaons, protons, and $\Lambda$) in small collision systems were extended to a wider transverse momentum ($p_{\mathrm{T}}$) range of up to 8 GeV/$c$ for the first time. The string- melting version of the AMPT model provides a good description of the measured $p_{\mathrm{T}}$-differential $v_{2}$ of the mesons but exhibits a slight deviation from the baryon $v_{2}$. In addition, we observed the features of mass ordering at low $p_{\mathrm{T}}$ and the approximate number of constituent quarks (NCQ) scaled at intermediate $p_{\mathrm{T}}$. Moreover, we demonstrate that hadronic rescattering does not have a significant impact on $v_{2}$ in p--Pb collisions for different centrality selections, whereas partonic scattering dominates in generating the elliptic anisotropy of the final particles. This study provides further insight into the origin of collective-like behavior in small collision systems and has referential value for future measurements of azimuthal anisotropy.
\end{abstract}


\maketitle

\section{Introduction}
The main goal of heavy-ion collisions at ultrarelativistic energies is to explore the deconfined state of strongly  interacting matter created at a high energy density and temperature, known as the gluon plasma (QGP)~\cite{Shuryak:1978ij,Shuryak:1980tp}. An important observation for investigating the transport properties of the QGP is anisotropic flow~\cite{Ollitrault:1992bk,Voloshin:2009fd}, which is quantified by the flow harmonic coefficients $v_{n}$ obtained from the Fourier expansion of the azimuthal distribution of the produced particles~\cite{Voloshin:1994mz,Poskanzer:1998yz}: 
\begin{equation}
\frac{dN}{d\varphi} \propto 1 + 2 \sum\limits_{n=1}^{\infty}v_{n}\mathrm{cos}[n(\varphi-\Psi_{n})],
\label{eq: def}
\end{equation}
where $\varphi$ is the azimuthal angle of the final-state particle angle and $\Psi_{n}$ is the symmetry plane angle in the collision for the $n$-th harmonic~\cite{Alver:2010gr,Alver:2010dn}. The second-order coefficient $v_{2}$, referred to as the elliptic flow, is derived from the initial state spatial anisotropy of the almond-shaped collision overlap region that is propagated to the final state momentum space. The magnitude of the elliptic flow is sensitive to the fundamental transport properties of the fireball, such as the temperature-dependent equation of state and the ratio of shear viscosity to entropy density ($\eta/s$)~\cite{Qin:2010pf,Teaney:2010vd}. 

Over the past few decades, various measurements of elliptic flow in heavy-ion collisions performed at the relativistic heavy-ion collider (RHIC)~\cite{BRAHMS:2004adc,PHENIX:2004vcz,PHOBOS:2004zne,STAR:2005gfr} and the Large Hadron Collider (LHC)~\cite{ALICE:2011ab,ALICE:2018rtz,ATLAS:2012at,CMS:2013wjq} have helped build a full paradigm of the strongly coupled QGP. Comprehensive measurements of $p_{\rm T}$-differential elliptic flow of the identified particles were conducted by the ALICE Collaboration~\cite{ALICE:2022zks,ALICE:2021ibz}. The observed mass-ordering effect (i.e., heavier particles have a smaller elliptic flow than lighter particles at the same $p_{\rm T}$) at low $p_{\rm T}$ is well described by hydrodynamic calculations and is attributed to the radial expansion of the QGP~\cite{Ma:Flow}. At intermediate $p_{\rm T}$, the grouping of $v_{2}$ of mesons and baryons was observed, with mesons exhibiting less $v_{2}$ than baryons. These behaviors can be explained by the hypothesis that baryons and mesons have different production mechanisms through quark coalescence, which has been further investigated using constituent quark (NCQ) scaling~\cite{Molnar:2003ff,Lin:2003jy,Wang:2022det,Yan:2006bx}. Interestingly, such flow-like phenomena have been observed in small-collision systems. Long-range double-ridge structures were first measured in high-multiplicity pp and p--Pb collisions by the ALICE, ATLAS, and CMS collaborations~\cite{CMS:2012qk,ALICE:2012eyl,ATLAS:2012cix}. The measurement of elliptic and triangular azimuthal anisotropies in central $^{3}\mathrm{He}$+Au, $d$+Au, and $p$+Au collisions performed by the STAR Collaboration~\cite{STAR:2022pfn} suggests that sub-nucleon fluctuations also play an important role in influencing the flow coefficients in these small collision systems. In addition, these measurements were extended to the identified particles associated with the discovery of a significant positive $v_{2}$~\cite{ALICE:2013snk,CMS:2018loe}. The observed particle-mass dependence of $v_{2}$ is similar to that measured in heavy-ion collisions~\cite{ALICE:2013snk}; however, the origin of such collective-like behavior remains unclear. Several theoretical explanations relying on either the initial state or final state effects have been proposed to understand the origin of azimuthal anisotropies in small systems. Studies that extend hydrodynamics from large to small systems based on final-state effects can well describe $v_{2}$ of soft hadrons~\cite{Bozek:2011if,Bozek:2012gr,Nagle:2018nvi,Nijs:2020roc,Nijs:2020ors}; however, they are based on the strong assumption that there is sufficient scattering among constituents in small systems. Hydrodynamics combined with the linearized Boltzmann transport (LBT) model can also describe the identified particle $v_{2}$ in a high-multiplicity small-collision system at an intermediate $p_{\mathrm{T}}$~\cite{Zhao:2020wcd}. Color-Glass Condensate (CGC)-based models and IP-Glasma models that consider the effect of momentum correlations in the initial state can quantitatively describe some features of collectivity in p--Pb collisions~\cite{Dusling:2013oia,Dusling:2012cg}, but without clear conclusions, particularly regarding the dependence on collision systems and rapidity. 

In addition, an approach called parton escape shows that few scatterings can also create sufficient azimuthal anisotropies, which have been investigated using multiphase transport (AMPT)~\cite{Lin:2004en,He:2015hfa}. The $v_{2}$ values of light hadrons measured in p--Pb collisions are well described in AMPT, where the contribution of anisotropic parton escape rather than hydrodynamics plays an important role~\cite{He:2015hfa}. In this study, we extend the AMPT calculations of the $p_{\rm T}$-differential $v_{2}$ for identified particles ($\pi^{\pm}$, $\mathrm{K}^{\pm}$, p($\bar{\mathrm{p}})$, $\Lambda(\bar{\Lambda})$) to higher $p_{\rm T}$ region in p--Pb collisions at 5.02 TeV, in order to systematically test the mass-ordering effect and baryon-meson grouping at low- and intermediate-$p_{\mathrm{T}}$, respectively. We also investigate how the key mechanisms implemented in AMPT, such as the parton cascade and hadronic rescattering, affect elliptic anisotropy in small collision systems. In addition, various nonflow subtraction methods with different sensitivities to jet-like correlations were studied. 

\section{Model and Methodology}
\subsection*{A multiphase transport model}
The string-melting version of the AMPT model (v2.26t9b, available online)~\cite{Lin:2004en,Lin:2021mdn} was employed in this study to calculate $v_{2}$ of the final-state particles in high-multiplicity p--Pb at 5.02 TeV. The AMPT model includes four main processes: initial conditions, partial scattering, hadronisation, and hadronic interactions. The initial conditions are generated from the heavy ion jet interaction generator (HIJING) model~\cite{Wang:1991hta,Gyulassy:1994ew}, where minijet partons and soft-excited strings are produced and then converted to primordial hadrons based on Lund fragmentation. Under the string-melting mechanism, primordial hadrons are converted into partons, a process determined by their flavor and spin structures. Elastic scattering between the partons was simulated using Zhang's parton cascade (ZPC) model~\cite{Zhang:1997ej}, which includes two-body scattering with a cross-section described by the following simplified equation:
\begin{equation}
\sigma_{gg}\approx\frac{9\pi\alpha_{s}^{2}}{2\mu^{2}}.
\end{equation}
In this study, the strong coupling constant $\alpha_{s}$ was set to 0.33, and the Debye screening mass $\mu$ = 2.2814~$\mathrm{fm}^{-1}$, resulting in a total parton scattering cross section of $\sigma=$ 3 mb. To isolate the effect of partonic scattering, $\sigma$ is adjusted to be close to 0 by increasing $\mu$ (see set "w/o parton scat." in Tab.~\ref{tab:Para}). Once the partonic interaction ceases, hadronization with a quark coalescence model is implemented to combine the nearest two (or three) quarks into mesons (or baryons)~\cite{Lin:2004en}. The formed hadrons enter the subsequent hadronic rescattering process using a relativistic transport (ART) model~\cite{Li:1995pra}, in which both elastic and inelastic scattering are considered for baryon-baryon, baryon-meson, and meson-meson interactions. The hadronic interaction time was set by default to $t_{max} = 30$ fm/c. Alternatively, $t_{max}$ is set to 0.4 fm/c to effectively turn off the hadron scattering process while still considering the resonance decay~\cite{Zheng:2016iia}(see set "w/o hadron scat." in Tab.~\ref{tab:Para}). In addition, the random orientation of the reaction plane was turned on and the shadowing effect was considered in this analysis.
\begin{table}[htbp]
\centering
  \caption{Details of three configurations}
     \label{tab:Para}
  \begin{tabular}{c|cc}
    \hline\hline\noalign{\smallskip}
     Description&~~~~~~~~~~~~$\sigma$(mb)~~~~~~~~~~~~~~~~~& $t_{\mathrm{max}}$(fm/c) \\
    \noalign{\smallskip}\hline\noalign{\smallskip}
    w/ all & ~~~~~~~~3~~~~~~~~~~~~~~~& 30 \\
    w/o parton scat.  & ~~~~~~~~$\sim$0~~~~~~~~~~~~~~~& 30 \\
    w/o hadron scat.  & ~~~~~~~~3~~~~~~~~~~~~~~~& 0.4\\
    \noalign{\smallskip}\hline
   \end{tabular}
\end{table}

\subsection*{Two-particle correlation and nonflow subtraction}
The two-particle correlation (2PC) method is widely used to extract the flow signal in small collision systems because it can suppress the non-flow contribution from long-range jet correlations~\cite{Ma:1995zzc,CMS:2012qk,ALICE:2012eyl,ATLAS:2012cix,ALICE:2013snk}. Similar to Eq.~\ref{eq: def}, the azimuthal correlation between two emission particles can be represented by $N^{\mathrm{pairs}}$ pairs of emitted particles (labeled as $C(\Delta \varphi)$) as a function of the relative angle $\Delta \varphi=\varphi^{a}-\varphi^{b}$ between particles $a$ and $b$ and expanded in the Fourier series as follows: 
\begin{equation}
\begin{aligned}
C(\Delta\varphi) = \frac{\mathrm{d}N^{\mathrm{pair}}}{d\Delta \varphi} \propto 1 + 2 \sum_{n=1}^{\infty}V_{n\Delta}(\it{p}_{\rm T}^{a},\it{p}_{\rm T}^{b})\mathrm{cos}[n(\Delta\varphi)],
\end{aligned}
\label{eq: Fourier 2PC} 
\end{equation}
where $V_{n\Delta}$ refers to the two-particle $n$-th order harmonic. In a pure hydrodynamic scenario, because particle emissions are independent, $V_{n\Delta}(\it{p}_{\rm T}^{a},\it{p}_{\rm T}^{b})$ can be factorized into the product of a single-particle flow $v_{n}^{a}$ and $v_{n}^{b}$:
\begin{equation}
\begin{aligned}
V_{n\Delta}(p_{\mathrm{T}}^{a},p_{\mathrm{T}}^{b}) = v_{n}(p_{\mathrm{T}}^{a})v_{n}(p_{\mathrm{T}}^{b}).
\end{aligned}
\label{eq: Factorization} 
\end{equation}
Based on the factorization assumption, $v_{n}$ of a single particle $a$ can be obtained using the 3$\times$2PC method, which was recently proposed by the PHENIX Collaboration~\cite{PHENIX:2022nht}. This requires the formation of two-particle correlations between three groups of particles (labeled $a$, $b$ and $c$) and the extraction of the flow coefficients for three combinations:  
\begin{equation}
\begin{aligned}
v_{n}(p_{\mathrm{T}}^{a}) = \sqrt{\frac{V_{n\Delta}(p_{\mathrm{T}}^{a},p_{\mathrm{T}}^{b})V_{n\Delta}(p_{\mathrm{T}}^{a},p_{\mathrm{T}}^{c})}{V_{n\Delta}(p_{\mathrm{T}}^{b},p_{\mathrm{T}}^{c})}}.
\end{aligned}
\label{eq: Fourier 3x2PC} 
\end{equation}
In small-collision systems, two main types of nonflow contributions to the flow signal are the near-side jet and away-side jet (recoil jet) correlations. The former can be effectively removed by introducing a large rapidity gap between the trigger and associated particles during the construction of the correlations. Several methods have been developed to subtract the latter ~\cite{Lim:2019cys}. A traditional approach is to directly subtract the correlation function distribution obtained from low-multiplicity events~\cite{ALICE:2012eyl,ALICE:2013snk} from that obtained from high-multiplicity events. This method assumes that the yield and shape of dijets are identical for both collision types as follows:
\begin{equation}
\begin{aligned}
C^{\mathrm{HM}}(\Delta\varphi) - C^{\mathrm{LM}}(\Delta\varphi) &\propto 1 + 2 \sum_{n=1}^{\infty}V_{n\Delta}\mathrm{cos}[n(\Delta\varphi)] \\
&= a_{0} + 2 \sum_{n=1}^{\infty}a_{n}\mathrm{cos}[n(\Delta\varphi)],
\end{aligned}
\label{eq: peripheral subtraction}
\end{equation}
where $C^{\mathrm{LM}}(\Delta\varphi)$ and $C^{\mathrm{HM}}(\Delta\varphi)$ represent the correlation function distributions obtained for low- and high-multiplicity events, respectively. This method relies on the "zero yield at minimum" (ZYAM) hypothesis~\cite{ALICE:2012eyl,ALICE:2013snk} that a flat combinatoric component should be subtracted from the correlation function in low-multiplicity events. Therefore, the fit parameter $a_{2}$ is the absolute modulation in the subtracted correlation function distribution and characterizes the modulation relative to a baseline, assuming that such a modulation is not present in the low-multiplicity class below the baseline. In this case, the flow coefficient $V_{n\Delta}$ is calculated as
\begin{equation}
\begin{aligned}
V_{n\Delta} = a_{n}/(a_{0}+b),
\end{aligned}
\label{eq: calculation of delta V2}
\end{equation}
where $b$ is the baseline, estimated using the minimum correlation function for low-multiplicity events. However, the measurement of jet-like correlations in p--Pb collisions indicates that the dependence of the dijet yield on the particle multiplicity cannot be ignored. In this case, a new template fit method was developed by the ATLAS collaboration~\cite{ATLAS:2015hzw}, where the correlation function distribution obtained in high-multiplicity events is assumed to result from the superposition of the distribution obtained in low-multiplicity events scaled up by a multiplicative factor $F$ and a constant modulated by $\mathrm{cos}(n\Delta\varphi)$ for $n > 1$, as shown in
\begin{equation}
\begin{aligned}
C(\Delta\varphi) = FC^{\mathrm{LM}}(\Delta\varphi) + G(1 + 2 \sum_{n=1}^{3}V_{n\Delta}\mathrm{cos}(n\Delta\varphi)),
\end{aligned}
\label{eq: template fit}
\end{equation}
where $G$ denotes the normalization factor that maintains the integral of $C(\Delta\varphi)$ equal to $C^{\mathrm{HM}}(\Delta\varphi)$. Furthermore, an improved template fitting method~\cite{ATLAS:2018ngv} developed in recent years was tested. It applies a correction procedure to the default template fit method by considering the multiplicity dependence of the remaining ridge in low-multiplicity events, as shown in
\begin{equation}
\begin{aligned}
V_{n\Delta} = V_{n\Delta}(\mathrm{tmp}) - \frac{FG^{\mathrm{LM}}}{G^{\mathrm{HM}}}(V_{n\Delta}^{2}(\mathrm{tmp}) - V_{n\Delta}^{2}(\mathrm{LM})),
\end{aligned}
\label{eq: improve template fit}
\end{equation}
where $V_{n\Delta}(\mathrm{tmp})$ and $V_{n\Delta}^{2}(\mathrm{LM})$ are obtained by using the default template method for high- and low-multiplicity events. All these nonflow subtraction methods are implemented in this study, and their different sensitivities to nonflow effect are also discussed.  

\section{Analysis procedures}
To directly  compare the AMPT calculations with the results from ALICE, we focused on the particles within the pseudorapidity range $|\eta| < 0.8$,  aligning with the TPC acceptance in ALICE~\cite{ALICE:2014sbx}. In the 3$\times$2PC method, long-range correlations were constructed between the charged particles at mid-rapidity, forward rapidity ($2.5 < \eta < 4$), and backward rapidity ($-4 < \eta < -2.5$), that is, the central-forward correlation ($-4.8 <\Delta\eta< -1.7$), central-backward correlation ($1.7 <\Delta\eta< 4.8$), and backward-forward correlations ($-8 <\Delta\eta< -5$). In addition, the centrality classes are defined by counting the charged particles in the acceptance of the V0A detector~\cite{ALICE:2014sbx}, that is, $2.8<\eta<5.1$. 

The correlation function distribution $C(\Delta\varphi)$ was obtained by correcting the number of particle pairs in the same events normalized to the number of trigger particles $N_{\mathrm{trig}}$ by using an event-mixing technique:
\begin{equation}
\begin{aligned}
C(\Delta\varphi, \Delta\eta) = \frac{1}{N_{\mathrm{trig}}}\frac{\mathrm{d}^{2}N_{\mathrm{pairs}}}{\mathrm{d}\Delta\eta \mathrm{d}\Delta\varphi} = \frac{S({\Delta \varphi, \Delta \eta})}{B({\Delta \varphi, \Delta \eta})},
\end{aligned}
\label{eq: per-trigger yield after Mixed 2PC}
\end{equation}
where $S({\Delta \varphi, \Delta \eta}) = \frac{1}{N_{\mathrm{trig}}}\frac{\mathrm{d}^{2}N_{\mathrm{same}}}{\mathrm{d}\Delta\eta \mathrm{d}\Delta\varphi}$ is the correlation function in same events and $B({\Delta \varphi, \Delta \eta}) = \alpha\frac{\mathrm{d}^{2}N_{\mathrm{mixed}}}{\mathrm{d}\Delta\eta \mathrm{d}\Delta\varphi}$ is the associated yield as a function of $\Delta \varphi$ and $\Delta \varphi$ in mixed events. Factor $\alpha$ is used to normalize $B({\Delta \varphi, \Delta \eta})$ to unity in the $\Delta\eta$ region of the maximal pair acceptance. The obtained 2-D correlation function $C(\Delta\varphi, \Delta\eta)$ is projected onto $\Delta\varphi$ axis, and we follow the nonflow subtraction procedures and factorization, as discussed in Eq. ~\ref{eq: Fourier 3x2PC}--Eq.~\ref{eq: improve template fit}, $v_{2}$ of the charged particles at $|\eta| < 0.8$ can be calculated.

\begin{figure}[!hbt]
\begin{center}
\includegraphics[width=.9\columnwidth]{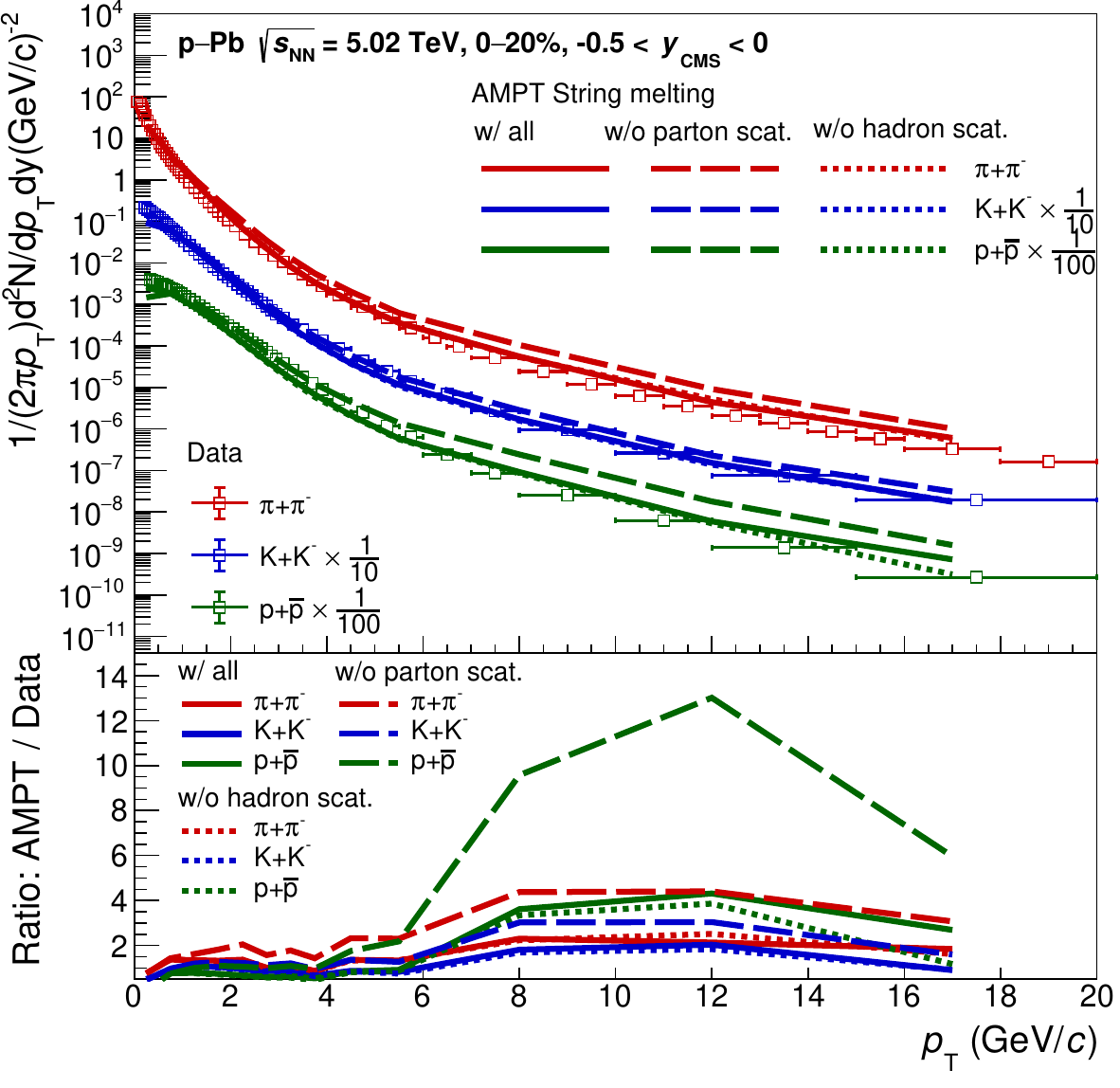}
\caption{(Color online) The $p_{\rm T}$ distribution of pions, kaons, and protons in 0--20\% high-multiplicity p--Pb collisions at $\sqrt{s_{\rm NN}}$ = 5.02 TeV, obtained from AMPT model calculations, is compared to ALICE measurement~\cite{ALICE:2016dei}. The results in AMPT without hadronic scattering and partonic scattering are also presented.}
\label{Fig: pT spectrum}
\end{center}
\end{figure}

\begin{figure*}[!hbt]
\begin{center}
\includegraphics[width=.9\columnwidth]{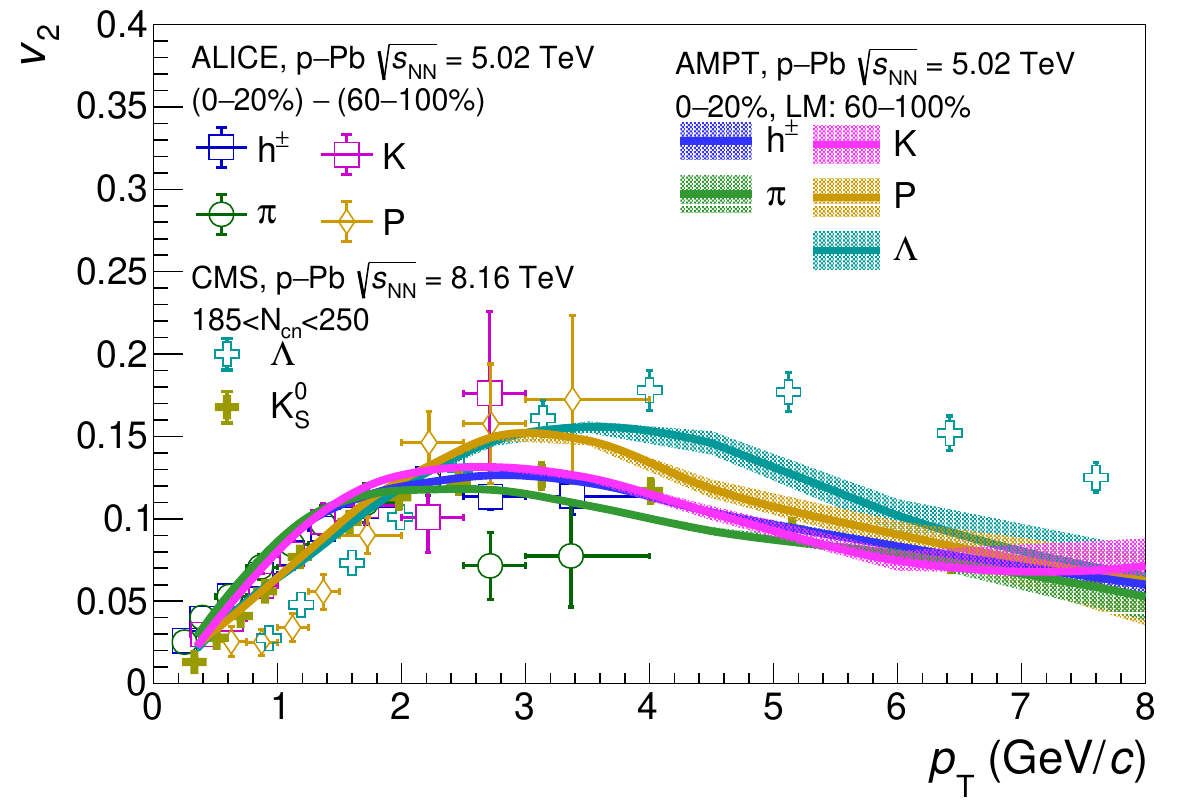}
\includegraphics[width=.9\columnwidth]{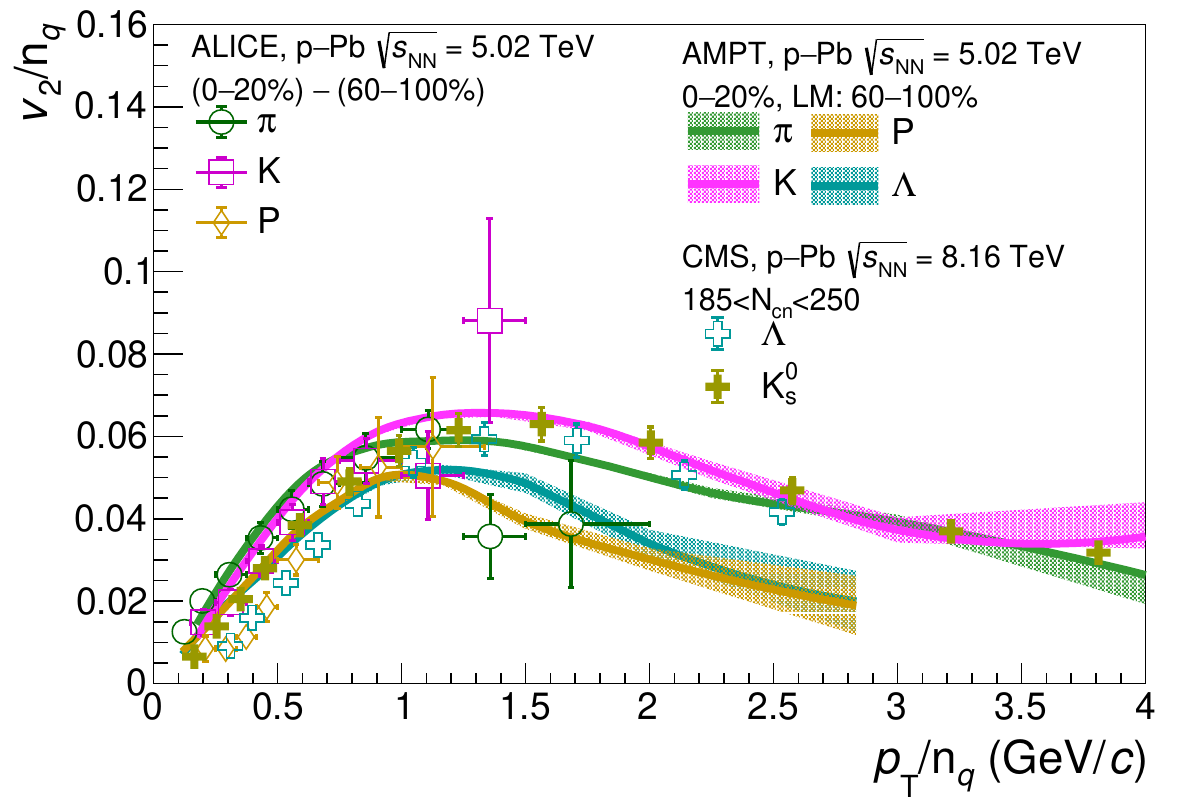}
\caption{(Color online) Left: the $v_{2}$ as a function of $p_{\rm T}$ in 0--20\% high-multiplicity p--Pb collisions at $\sqrt{s_{\rm NN}}$ = 5.02 TeV, obtained from default AMPT model calculations with 3$\times$2PC method, is compared to ALICE and CMS measurement~\cite{ALICE:2013snk,CMS:2018loe}. Right: the $p_{\rm T}$-differential $v_{2}$ scaled by the number of constituent quark ($n_{q}$).}
\label{Fig: v2 with FB}
\end{center}
\end{figure*}

\begin{figure*}[!hbt]
\begin{center}
\includegraphics[width=.9\columnwidth]{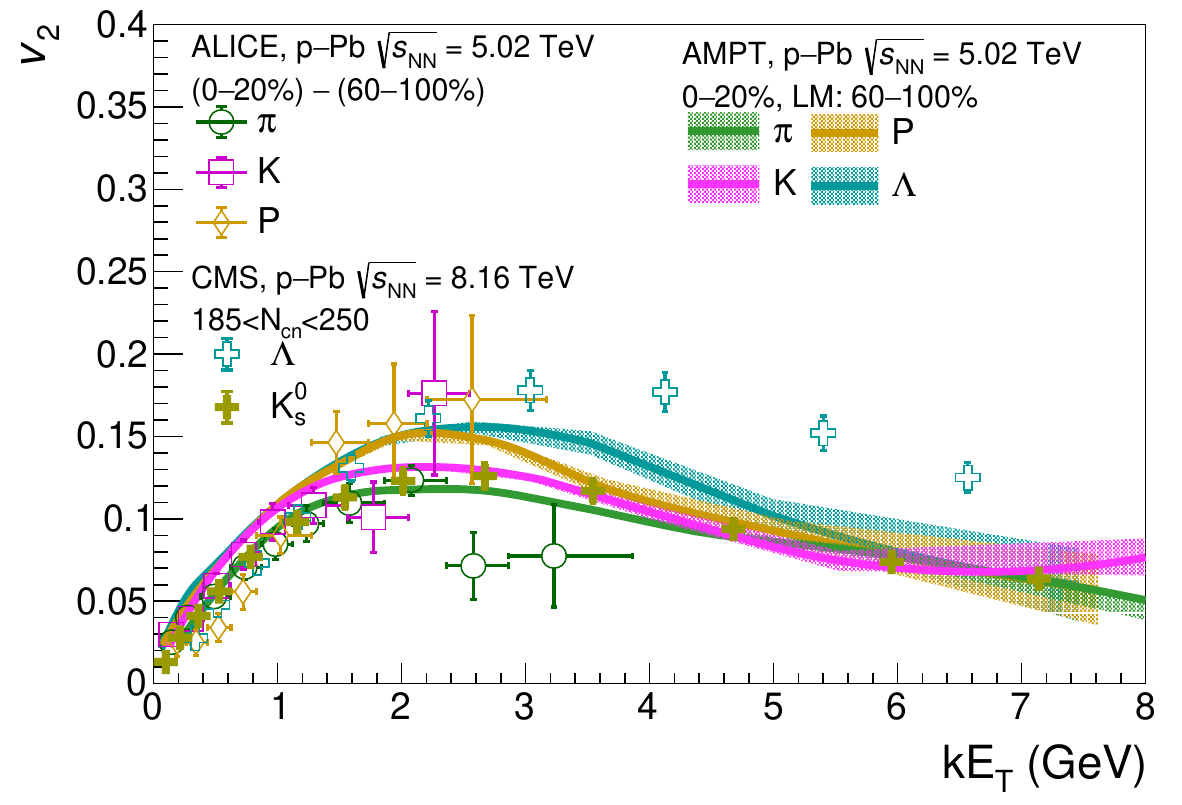}
\includegraphics[width=.9\columnwidth]{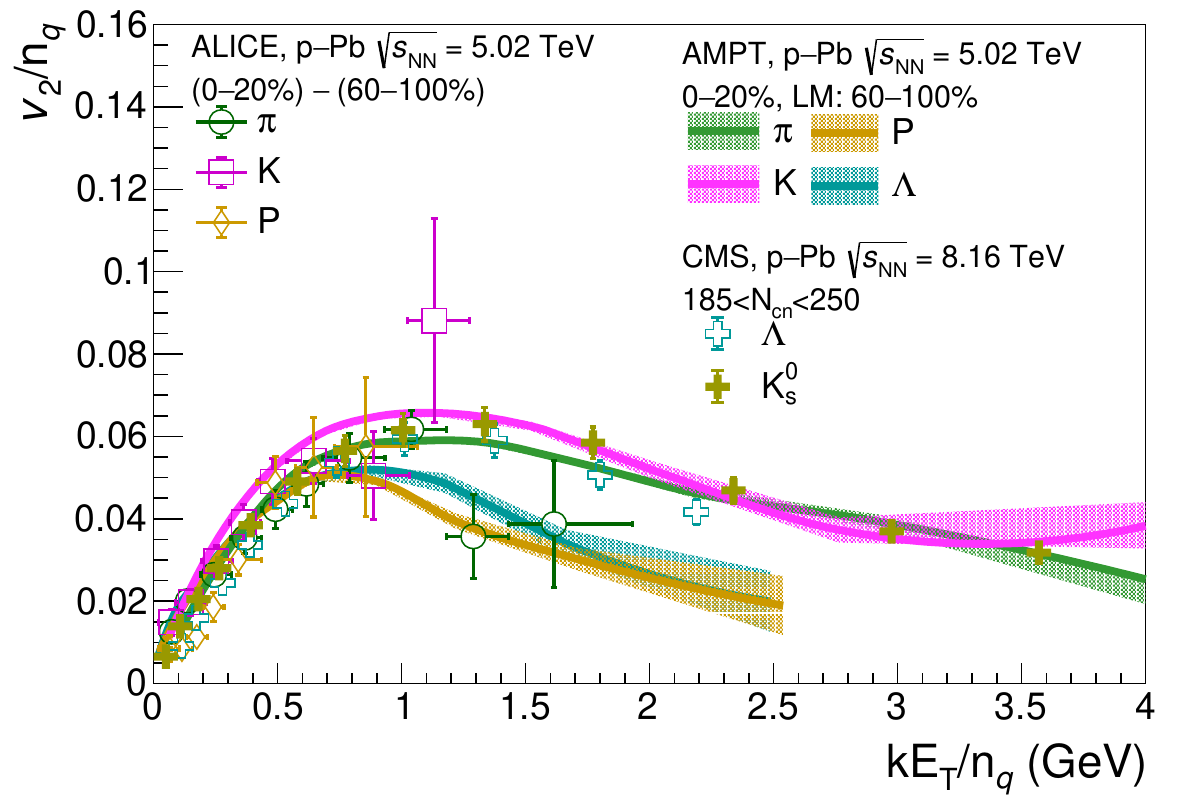}
\caption{(Color online) Left: the $v_{2}$ as a function of transverse kinetic energy (k$\mathrm{E}_{\mathrm{T}}$) in 0--20\% high-multiplicity p--Pb collisions at $\sqrt{s_{\rm NN}}$ = 5.02 TeV, obtained from default AMPT model calculations with 3$\times$2PC method, is compared to ALICE and CMS measurement~\cite{ALICE:2013snk,CMS:2018loe}. Right: the k$\mathrm{E}_{\rm T}$-differential $v_{2}$ scaled by the number of constituent quark ($n_{q}$).}
\label{Fig: KET v2 with FB}
\end{center}
\end{figure*}

\section{Results and Discussions}
We first investigated the $p_{\rm T}$ spectrum of the identified particles before performing the flow analysis. Figure~\ref{Fig: pT spectrum} illustrates the $p_{\rm T}$ distribution of proton, pion, and kaon in 0--20\% high-multiplicity p--Pb collisions at $\sqrt{s_{\rm NN}}$ = 5.02 TeV, which are obtained from AMPT with three different sets of configurations listed in Tab. ~\ref{tab:Para} and ALICE experimental data~\cite{ALICE:2016dei}. The AMPT results, both with and without hadronic rescattering, are consistent. This behavior differs from previous findings in heavy-ion collisions, where the hadronic interaction significantly reduces the particle yield ~\cite{Lin:2000cx}. The spectrum obtained in the AMPT without considering the parton cascade process is enhanced compared to that obtained with partonic scattering, and this enhancement is more significant at a high $p_{\rm T}$. This outcome is expected because partons experience energy loss during the parton cascade, which reduces the production of final-state particles. In addition, the ratios of the $p_{\rm T}$ spectra obtained from the AMPT calculations and data are shown. The AMPT model calculation reproduces the particle yields well at low and intermediate $p_{\rm T}$ values when both partonic and hadronic scattering are included; however, it overestimates the high $p_{\rm T}$ data because parton-parton inelastic collisions and, subsequently, hard parton fragmentation are absent in the model~\cite{Lin:2021mdn}.

Figure~\ref{Fig: v2 with FB} (left) shows the $v_{2}$ of pions, kaons, protons and $\Lambda$ as a function of $p_{\rm T}$ in 0--20\% high-multiplicity p--Pb collisions at $\sqrt{s_{\rm NN}}$ = 5.02 TeV, obtained in AMPT calculations with 3$\times$2PC method. A comparison with the ALICE measurement for $v_{2}$ of charged hadrons, pions, kaons, and protons~\cite{ALICE:2013snk} and the CMS measurement for $v_{2}$ of $K_{s}^{0}$ and $\Lambda$~\cite{CMS:2018loe} is also presented. The AMPT calculations applied the template fit method to suppress the away-side jet contribution and considered the ZYAM assumption to enable direct comparison with the observed data. The $v_{2}$ values of charged hadrons, pions, and kaons can be described well by AMPT calculations, but the $v_{2}$ values of baryons (protons and $\Lambda$) cannot be reproduced. In addition, the mass-ordering effect (i.e., the $v_{2}$ of baryons is lower than that of mesons) is reproduced for $p_{\rm T} <2$ GeV/$c$. Owing to the advanced flow extraction method, the calculations of $v_{2}$ were extended to the high-$p_{\rm T}$ region, up to 8 GeV/$c$ in the AMPT model for the first time. The $v_{2}$ values of protons and $\Lambda$ are consistent, and both of them are observed to have a higher value of $v_{2}$ value than that of the mesons for $2<p_{\rm T} <$ 7 GeV/$c$. The observed meson-baryon particle type grouping in heavy-ion collision flow measurements indicates collective behavior at the partonic level, leading to the coalescence of quarks into hadrons. The number of constituent quarks (NCQ) scaling techniques described in~\cite{Molnar:2003ff} can be used for further studies of this grouping. $v_{2}$ and $p_{\rm T}$ in Fig. ~\ref{Fig: v2 with FB} (left) are replaced by $v_{2}/n_{q}$ and $p_{\mathrm{T}}/\mathrm{n}_{q}$, where the $n_{q}$ is the number of constituent quark in mesons ($n_{q} = 2$) and baryons ($n_{q} = 3$), as shown in Fig. ~\ref{Fig: v2 with FB} (right). $v_{2}/n_{q}$ obtained from the data show approximate values at intermediate $p_{\rm T}$; however, the results calculated in AMPT cannot reproduce the scaling in $p_{\rm T}/\mathrm{n}_{q}>1$ GeV/$c$. In order to consider the observed mass hierarchy of $v_{2}$, we also plot the $v_{2}$ of identified particle as a function of the transverse kinetic energy k$\mathrm{E}_{\mathrm{T}}$ (k$\mathrm{E}_{\mathrm{T}} = m_{\mathrm{T}} - m_{0} = \sqrt{p_{\mathrm{T}}^{2}+m_{0}^{2}} - m_{0}$), and its NCQ scaling in Fig. ~\ref{Fig: KET v2 with FB} (left), and Fig. ~\ref{Fig: KET v2 with FB} (right)). All particle species showed a set of similar $v_{2}$ values after NCQ scaling in k$\mathrm{E}_{\mathrm{T}}/n_{q} < 1$ GeV, confirming that the quark degree of freedom in flowing matter can also be probed in the transport model. However, this NCQ scaling is violated for k$\mathrm{E}_{\mathrm{T}}/n_{q} > 1$ GeV. This may be attributed to the hadronization mechanism implemented in the AMPT model used in this study, where baryons are produced only after the formation of mesons by simply combining the three nearest partons, regardless of the relative momentum among the coalescing partons. This results in an underestimation of the baryon $v_{2}$ at intermediate $p_{\mathrm{T}}$ in this study. An improved coalescence model implemented in the newer AMPT~\cite{He:2017tla} introduced a new coalescence parameter to control the relative probability of a quark forming a baryon instead of a meson precisely. This improvement could have different NCQ scaling on $v_{2}$ but requires more systematic studies. Further studies on $v_{2}$ calculations in small collision systems with other improved hadronization mechanisms, for example, considering the Wigner function~\cite{Wang:2019xph} and hard parton fragmentation~\cite{Zhang:2022fum}, should be performed in the future. 

\begin{figure}[!hbt]
\begin{center}
\includegraphics[width=.9\columnwidth]{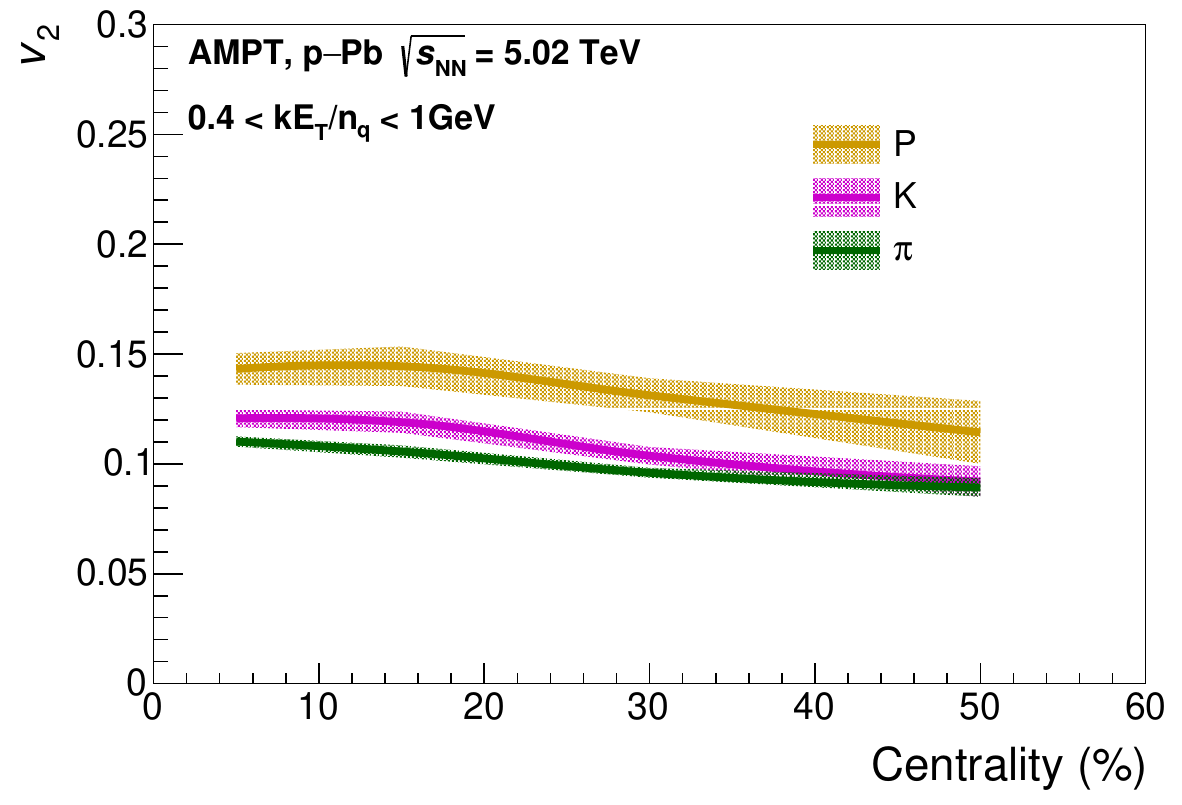}
\caption{(Color online) The integrated $v_{2}$ in 0.4$<\mathrm{kE}_{\mathrm{T}}/n_{q}<$1 GeV for pion, kaon and proton varying with the centrality.}
\label{Fig: v2 vs Cen}
\end{center}
\end{figure}

\begin{figure}[!hbt]
\begin{center}
\includegraphics[width=.9\columnwidth]{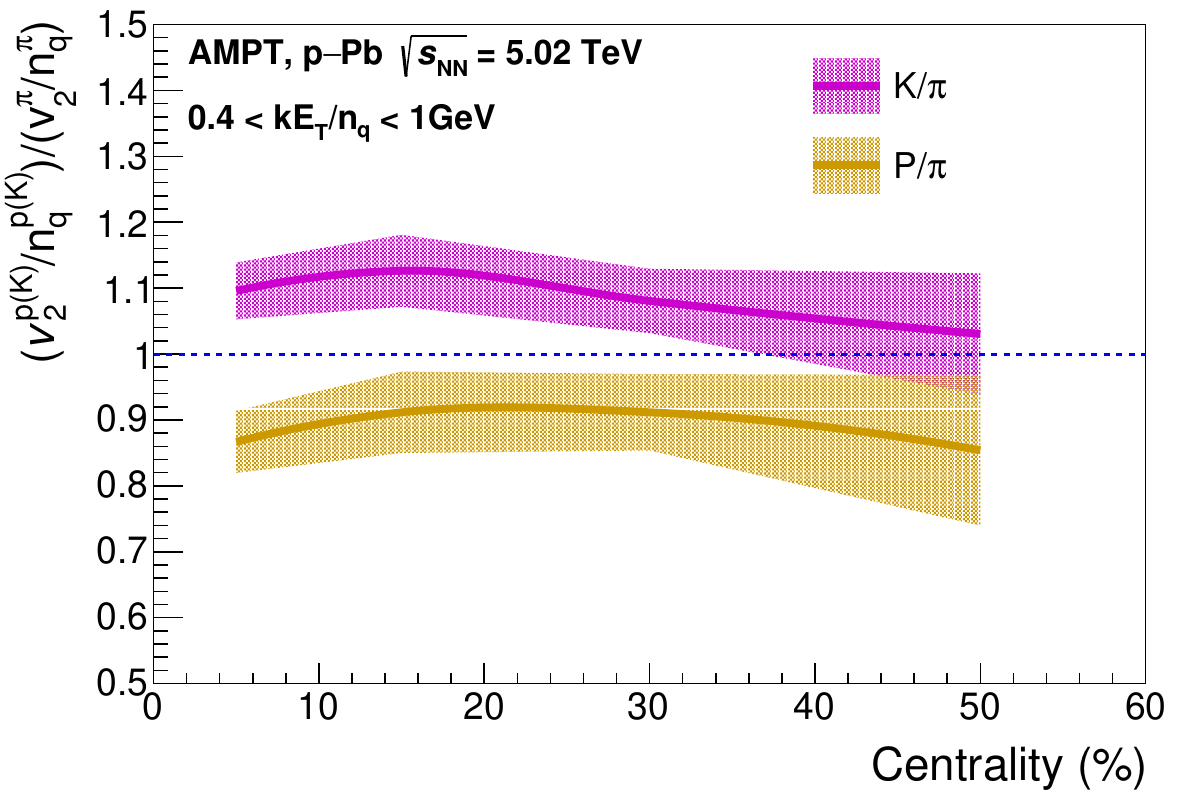}
\caption{(Color online) The ratio of integrated $v_{2}$ within 0.4$<\mathrm{kE}_{\mathrm{T}}/n_{q}<$1 GeV for proton over pion and kaon over pion varying with the centrality. The dash line represents the location of unity ratio.}
\label{Fig: v2 vs Cen ratio}
\end{center}
\end{figure}

We also extend our investigation to include a study of integrated $v_{2}$ within various centrality bins spanning the 0--60\% range. We focus on the region where the NCQ scaling criterion is satisfied, that is, for transverse kinetic energies per constituent quark ($\mathrm{kE}_{\mathrm{T}}/n_{q}$) ranging from 0.4 to 1 GeV. The non-flow contribution was estimated and subtracted within the 60--100\% centrality class by using the template fit method. As shown in Fig.~\ref{Fig: v2 vs Cen}, the $v_{2}$ values as a function of centrality exhibit a systematic decrease from central to peripheral collisions, reflecting the changing dynamic conditions and particle production mechanisms in different collision zones. Intriguingly, in the $v_{2}$ measurements, we observed a distinct mass-splitting phenomenon, with baryons and mesons exhibiting distinct elliptic flow patterns. Such a mass dependence in $v_{2}$ is similar to that in heavy-ion collisions at the LHC energies presented in a previous study~\cite{Zheng:2016iia}. This provides valuable insights into the collective behavior of different particle species within the evolving fireball created during these collisions.

Moreover, to gain a deeper understanding of the NCQ scaling properties, we explored the ratios of $\mathrm{n}_{q}$-scaled integrated $v_{2}$ values for protons relative to pions and kaons relative to pions as functions of centrality. The results are shown in Fig. ~\ref{Fig: v2 vs Cen ratio}. A notable trend is observed in these ratios: they tend to approach unity as the collisions become more peripheral. It indicates that the collective flow of particles in low-multiplicity events may be approaching a behavior that is closer to the expected scaling behavior based on the number of constituent quark.

\begin{figure}[hbt]
\begin{center}
\includegraphics[width=.9\columnwidth]{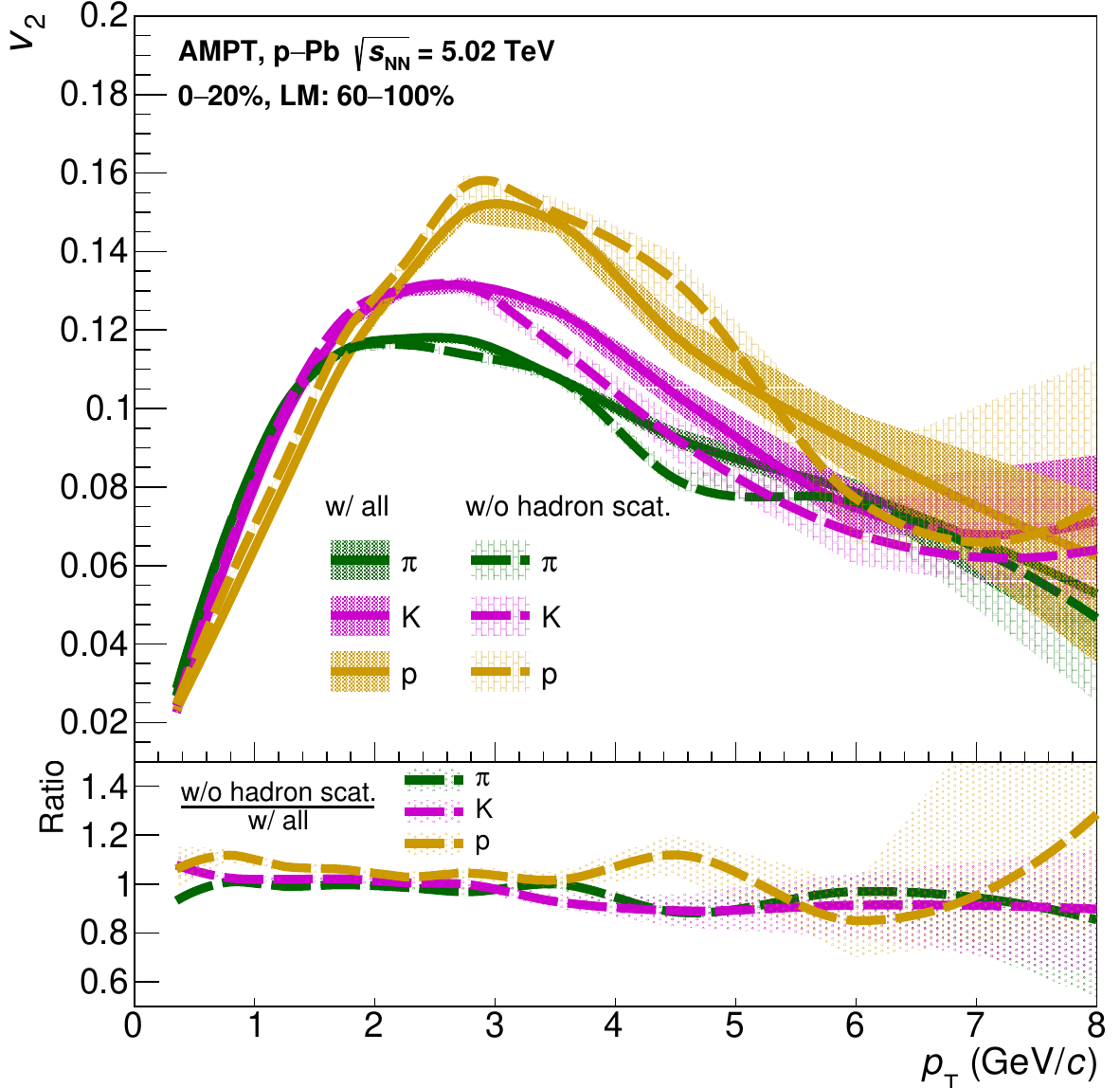}
\caption{(Color online) The $p_{\rm T}$-differential $v_{2}$ of pions, kaons, and protons calculated in AMPT model with and without considering hadronic scattering. The ratios of the two sets are also presented.}
\label{Fig: v2 with ART}
\end{center}
\end{figure}

\begin{figure}[hbt]
\begin{center}
\includegraphics[width=.9\columnwidth]{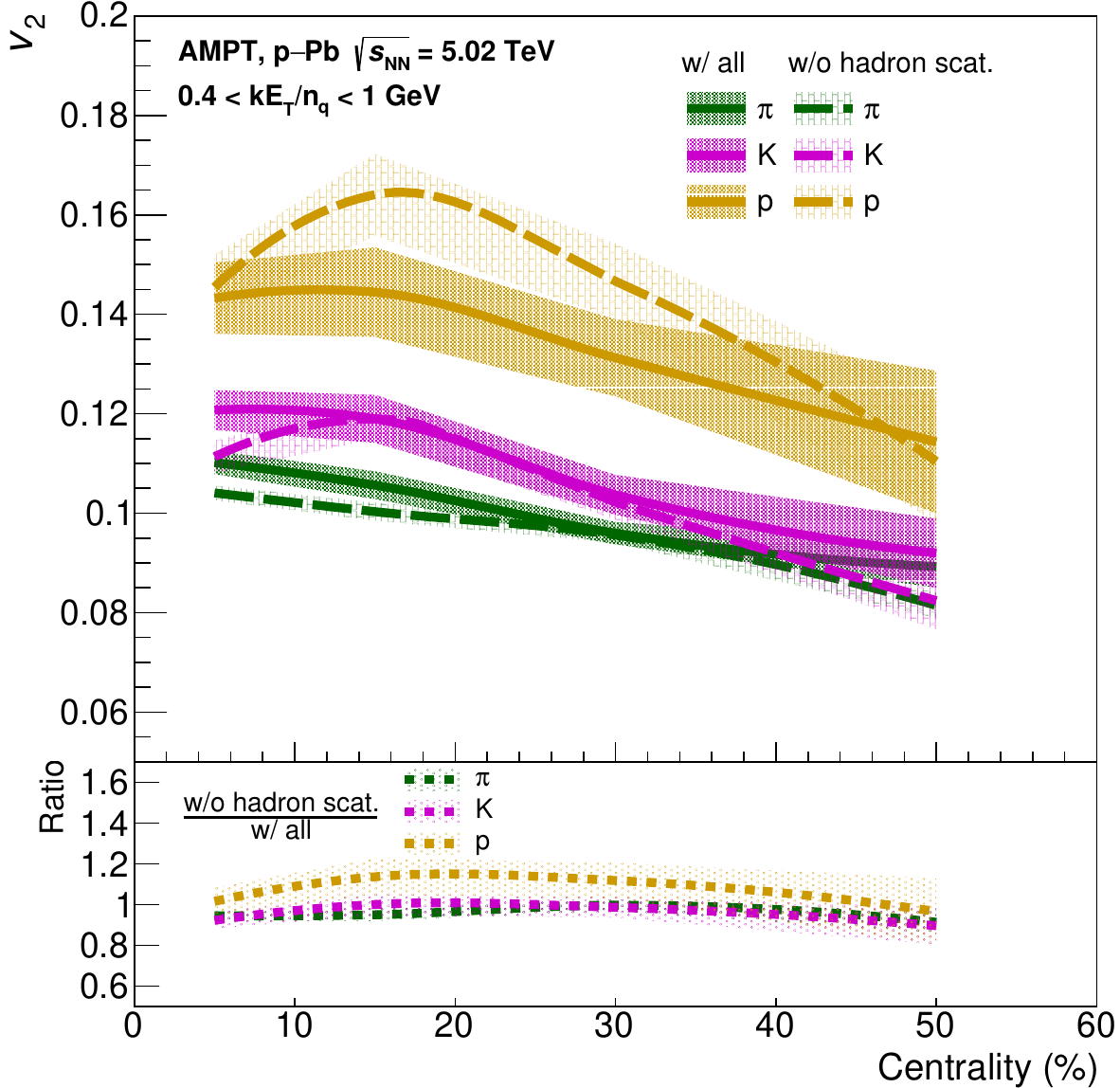}
\caption{(Color online) The integrated $v_{2}$ in 0.4$<\mathrm{kE}_{\mathrm{T}}/n_{q}<$1 GeV for pions, kaons, and protons calculated in AMPT model with and without considering hadronic scattering. The ratios of the two sets are also presented.}
\label{Fig: integrated v2 with ART}
\end{center}
\end{figure}

\begin{figure}[!hbt]
\begin{center}
\includegraphics[width=.9\columnwidth]{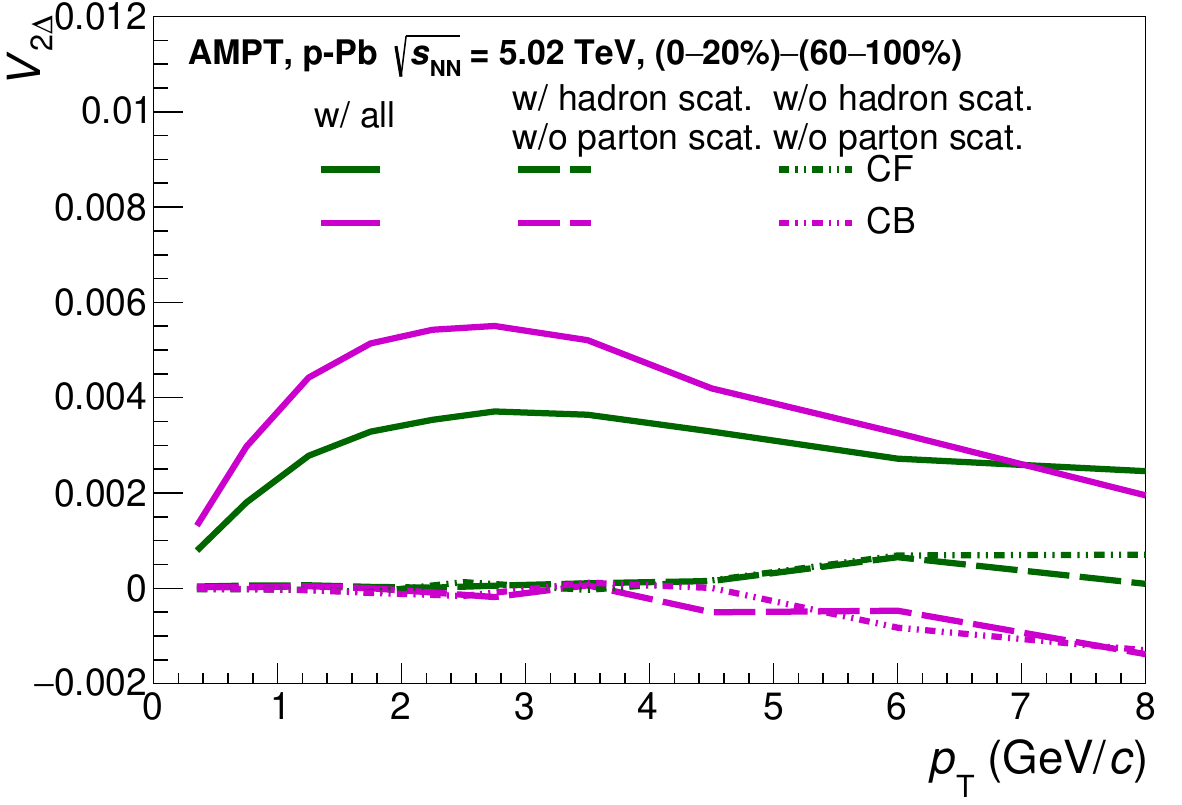}
\caption{(Color online) The $p_{\rm T}$-differential $V_{2\Delta}$ for central-forward (CF) and central-backward (CB) correlations calculated in AMPT model with and without considering partonic scattering.}
\label{Fig: v2 with ZPC}
\end{center}
\end{figure}

The effects of partonic and hadronic scattering on the elliptical anisotropy of the final-state particles were examined in this study. Figure~\ref{Fig: v2 with ART} shows the calculated $p_{\mathrm{T}}$-differential $v_{2}$ of pions, kaons, and protons in AMPT with and without considering hadronic rescattering process in 0—20\% high-multiplicity p—Pb collisions. The results show that the ratio of the $v_{2}$ values with and without hadronic rescattering is consistent with unity  for all particle species, indicating that the hadronic rescattering mechanism has almost no effect on $v_{2}$ in high-multiplicity p—Pb collisions. We also investigated the centrality dependence of the hadronic rescattering effects by calculating $p_{\mathrm{T}}$-integrated $v_{2}$ in several centrality bins between 0 and 60\%, as illustrated in Fig. ~\ref{Fig: integrated v2 with ART}. The results demonstrate that the influence of hadronic rescattering is independent of the centrality selection and has almost no impact on NCQ scaling in the range of 0.4$<\mathrm{kE}_{\mathrm{T}}/n_{q}<$1 GeV. 

On the other hand, when we set the parton scattering cross-section $\sigma$ to zero but maintain the hadronic scatterings, the $V_{2\Delta}$ of charged particles for the central-forward (CF) and central-backward (CB) correlations is almost zero, as shown in Fig. ~\ref{Fig: v2 with ZPC}. If both the partonic and hadronic scatterings are turned off, the results remain consistent with zero. This indicates that the elliptical anisotropy in high-multiplicity small-collision systems is mostly generated by parton scattering. Our conclusion is consistent with previous studies on the AMPT~\cite{He:2015hfa}, which suggested that the majority of elliptic anisotropies comes from the anisotropic escape probability of partons.

\begin{figure}[!hbt]
\begin{center}
\includegraphics[width=.9\columnwidth]{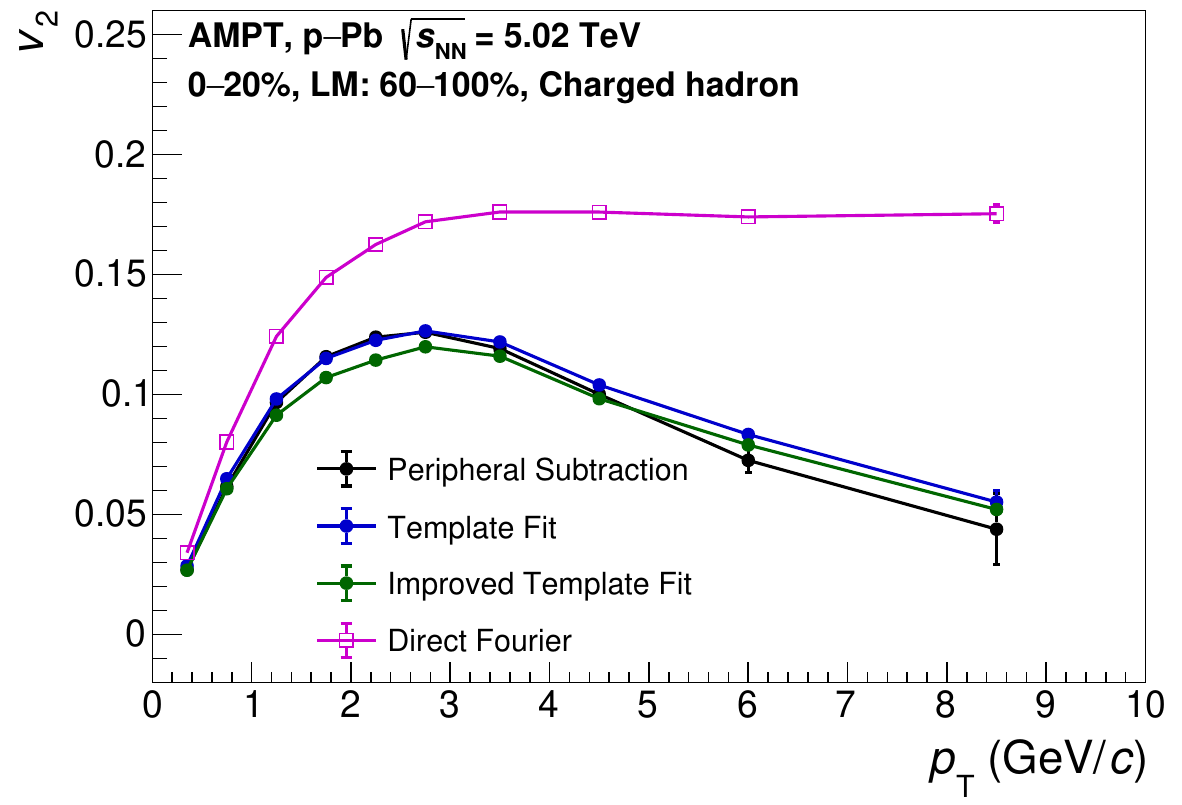}
\caption{(Color online) The $p_{\rm T}$-differential $v_{2}$ of charged hadrons calculated in AMPT with different nonflow subtraction methods.}
\label{Fig: v2 with nonflow test}
\end{center}
\end{figure}

Finally, different non-flow subtraction methods were investigated in this study. Figure~\ref{Fig: v2 with nonflow test} (left)  shows the $p_{\rm T}$-differential $v_{2}$ of the charged particles calculated using the 3$\times$2PC method in 0—20\% high-multiplicity p—Pb collisions. Several nonflow subtraction methods are implemented. To demonstrate how the nonflow contribution is removed, $v_{2}$ obtained with a direct Fourier transform of the $C(\Delta\varphi)$ correlation (as shown in Eq. ~\ref{eq: Fourier 2PC}). The results show significant suppression across all the subtraction methods, particularly at higher $p_{\rm T}$ values where jet correlations are dominant. The results obtained with peripheral subtraction and template fitting were consistent, indicating that the away-side jet contribution was automatically removed using the 3$\times$2PC method, even though the dependence of the jet correlation on multiplicity was not considered in the peripheral subtraction method. The $v_{2}$ calculated using the improved template fit method was slightly lower than that from the template fit, and it was similar to the features observed in the ATLAS measurement~\cite{ATLAS:2018ngv}. The same conclusions were drawn for the extraction of the identified particles (pions, kaons, protons, and $\Lambda$) $v_{2}$.

\section{Summary}
This study systematically investigated the elliptic anisotropy of identified particles (pions, kaons, protons, and $\Lambda$) in p—Pb collisions at 5.02 TeV using the AMPT model. We extended the calculation of $v_{2}$ to higher $p_{\rm T}$ regions, up to 8 GeV/$c$, using advanced nonflow subtraction techniques for the first time. We also examined the mass-ordering effect and baryon-meson grouping at low and intermediate $p_{\mathrm{T}}$, respectively. We argue that, with the approximate NCQ scaling of baryons and mesons, $v_{2}$ can be reproduced well at $\mathrm{kE}_{\mathrm{T}}/n_{q}<$1 GeV for several centrality bins. Furthermore, we demonstrate that parton interactions can simultaneously decrease the yield of light hadrons and generate significant $v_{2}$. However, hadronic rescatterings had little influence on the elliptical anisotropy of the final-state particles. Thus, these findings indicate that the nonequilibrium anisotropic parton escape mechanism coupled with the quark coalescence model can also reproduce the hydro-like behavior of the identified particles observed in small collision systems. Overall, this study provides new insights into the existence of partonic collectivity in small collision systems.

\section{Acknowledgement}
We would like to thank Guo-Liang Ma for the helpful discussions.

\end{document}